\documentclass[reprint,superscriptaddress,twocolumn,aps,pre,showpacs,floatfix]{revtex4}
\usepackage{graphicx,float}
\usepackage{amsbsy,amssymb,amsmath,bm,ulem}
\usepackage{enumerate}
\newcommand{\pd}[2]{\frac{\partial #1}{\partial #2}}

\begin{document}
\title{Accuracy of energy measurement and reversible operation of a microcanonical Szilard engine}
\author{Joakim Bergli}
\affiliation{Department of Physics, University of Oslo, P.O.Box 1048
  Blindern, N-0316 Oslo, Norway}

\begin{abstract}
  In a recent paper [Vaikuntanathan and Jarzynski, Phys. Rev. E {\bf
    83}, 061120 (2011)] a model was introduced whereby work could be
  extracted from a thermal bath by measuring the energy of a particle
  that was thermalized by the bath and manipulating the potential of
  the particle in the appropriate way, depending on the measurement
  outcome. If the extracted work is $W_1$ and the work $W_{\text{er}}$
  needed to be dissipated in order to erase the measured information
  in accordance with Landauer's principle, it was shown that $W_1\leq
  W_{\text{er}}$ in accordance with the second law of
  thermodynamics. Here we extend this work in two directions: First,
  we discuss how accurately the energy should be measured. By
  increasing the accuracy one can extract more work, but at the same
  time one obtains more information that has to be deleted. We discuss
  what are the appropriate ways of optimizing the balance between the
  two and find optimal solutions. Second, whenever $W_1$ is strictly
  less than $W_{\text{er}}$ it means that an irreversible step has
  been performed. We identify the irreversible step and propose a
  protocol that will achieve the same transition in a reversible way,
  increasing $W_1$ so that $W_1 = W_{\text{er}}$.

\end{abstract}
\maketitle

\section{Introduction}\label{sec:intro}

One of the various statements of the second law of thermodynamics is
the Kelvin-Planck formulation: No process is possible whose sole
result is the conversion of thermal energy to mechanical work. One
consequence of this is the following. Take a system which is initially
in thermal equilibrium, but then isolated from  the
environment. There is no way to reduce on  average  the energy
of the system by any cyclic variation of external parameters. If this
were possible, one could then reconnect the system with the thermal
bath and return it to the initial state with the only result that some
of the initial thermal energy was extracted as mechanical work in
contradiction with the second law. In fact, since the system is to be
isolated during the time when energy is to be extracted, this is a
statement about the possible time evolution of a dynamical
(Hamiltonian) system with a certain distribution of initial
conditions. Indeed, it can be directly proven from properties of
Hamiltonian systems
\cite{allahverdyan2002,campisi2008,jarzynski1997} and is
true not only for the canonical distribution of initial states, but
for any distribution function which is a monotonically decreasing
function of energy \cite{allahverdyan2002,campisi2008}.

Recently, the violation of this statement in the case of a
microcanonical ensemble of systems was discussed in several papers
\cite{sato2002,allahverdyan2002,marathe2010,vaikuntanathan2011}. That
is, if you know the initial energy of the system (but not the precise
initial state), you can find a cyclic variation of external parameters
such that the average energy of the system is reduced, and therefore
work on average extracted.  But a canonical ensemble can be
``converted'' to a microcanonical one by a measurement of the
energy. This idea was explored in Ref.  \cite{vaikuntanathan2011},
where a model is constructed which consists of a single particle in a
one-dimensional potential $U(q)$ (where $q$ is the position of the
particle). It is then shown that if you measure the energy you can
find a cyclic variation of the potential which reduces the energy of
the system as close to zero as you wish. The initial energy of the
particle is then delivered as work $W_1$ to the agent operating the
potential. An explicit protocol is given for the evolution of the
potential in the case where the initial potential is quartic,
$U(q)\sim q^4$, but the procedure is easily extended to other
potentials by adding a step that transforms the initial potential
adiabatically to a quartic one, whereby the ordering of the different
energy states is kept similarly to what is exploited in
\cite{vaikuntanathan2011}, and then performing the cyclic operation
they presented. They consider the following sequence of steps:

\begin{enumerate}[(1)]
\item The system is brought into contact and allowed to
equilibrate with a thermal reservoir at temperature $T$. The
reservoir is then removed.
\item The energy of the now-isolated system is measured.
\item The system is subjected to a cyclic protocol that reduces its
  kinetic energy close to zero, extracting on average the work $W_1$.
\end{enumerate}
This sequence can be repeated indefinitely, and thereby one has
constructed a device which converts thermal energy into mechanical
work, in seeming contradiction to the second law. The resolution if
the inconsistency is found in Landauer's principle
\cite{landauer1961}, which states that the erasure of information by
necessity results in dissipation of heat. That is, to erase the
information obtained when mesuring the energy of the system, and
restore the mesuring device to the initial state, one needs on average
an amount of work $W_{\text{er}}$ which is converted to thermal
energy. In \cite{vaikuntanathan2011} it is explicitly shown that we
have
\begin{equation}\label{work}
W_{\text{er}}\geq W_1.
\end{equation}
This means that to erase the information one
needs at least as much energy as one extracted from the thermal bath
by the operation of the device. 

The analysis presented in \cite{vaikuntanathan2011} shows how the
second law is not violated by such a device, but it leaves several
puzzles. In order to efficiently extract work from the system it is
necessary to know the energy accurately. But an accurate energy
measurement means a large amount of information. It seems that
$W_{\text{er}}\rightarrow\infty$ in the limit of very precise energy
measurements. At the same time, $W_1$ is bounded by the average energy
of the system at the time of the measurement. Is there some optimal
accuracy with which the energy should be measured in order to extract
as large a fraction of the energy as possible while still not having
to pay too much in deleting the information? And since the probability
of a certain energy depends on the energy, are there some regions in
energy where it is more important to make accurate mesurements? It is
also interesting to understand why in Eq.~\eqref{work} we sometimes
have an inequality, rather than strict equality. In other words, where
in the process is there an irreversible step which leads to a net
increase of the entropy? In this paper we will address these
questions.

The paper is organized as follows: In Sec. \ref{model} we present the
model and find the extracted work and measured information. Different
ways of optimizing the extracted work are discussed in
Sec. \ref{optimal}, and a protocol for reversibly completing the whole
operation in Sec. \ref{reversible}. A short summary and discussion is
given in Sec. \ref{summary}.

\section{Model}\label{model}

\begin{center}
\begin{figure}
\includegraphics[width=\linewidth]{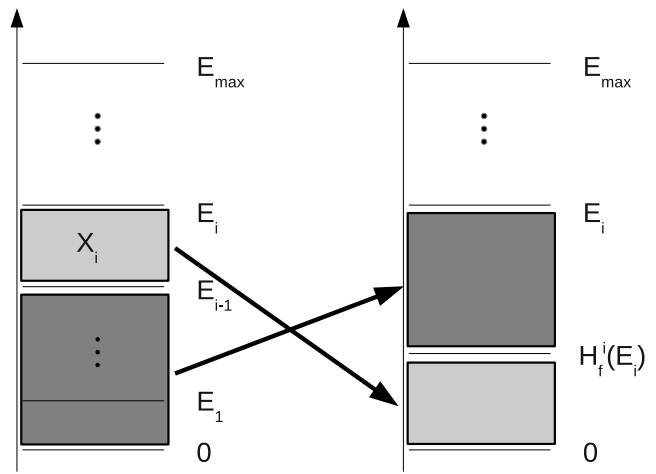}
\caption{\label{fig:energyExchange}The result of applying the protocol
  appropriate for an initial energy in the interval $X_i$. The
  manipulation protocol chosen will shift this interval to the lowest
  energies, and pushing those below up.}
\end{figure}
\end{center}

We use the model of Vaikuntanathan and Jarzynski~\cite{vaikuntanathan2011}.  
The system consists of a single particle
with coordinate $q$ and momentum $p$ in a potential $U(q)$. In
Ref. \cite{vaikuntanathan2011} they choose $U(q) \sim q^4$. The particle
is in thermal equilibrium at a certain temperature $T$. We assume that
we can mesure the energy of the particle to a certain accuracy. More
precicely, we define a number of energies $0<E_1\cdots<E_n<E_{\text{max}}$
between 0 and some maximal energy $E_{\text{max}}$ and we assume that we can
measure in which interval $X_i = [E_{i-1},E_i]$ the energy
lies. Depending on the outcome of the measurement, we choose an
appropriate manipulation protocol. The manipulation consists in
adiabatically modifying the shape of the potential through a closed
path in the space of potentials, returning it in the end to the
initial one, the exact protocol is given in
Ref.~\cite{vaikuntanathan2011}.  The end result is that the states
which initially were in the interval $X_i$ are shifted to the lowest
energies, see Fig. \ref{fig:energyExchange}.
Moreover, the ordering of the states inside the interval $X_i$ is
kept, so that a state with a lower initial energy will also have a
lower final energy. This means that we can find the final energy
$H_f^i(E)$ of a particle with initial energy $E$ when the protocol
appropriate for an energy in the interval $X_i$ is executed.  If
$g(E)$ is the density of states in the potential $U(q)$ we have
\begin{equation}\label{H0}
 \int_{E_{i-1}}^E \! dE\,  g(E) = \int_0^{H_f^i(E)} \!dE \, g(E)
\end{equation}
and $H_f^i(E)$ is found by solving this equation. In the case of a
one-dimensional system in a quadratic potential $U(q) \sim q^2$ the
density of states is a constant, $g(E) = g_0$, and we obtain a
particularly simple equation which gives
\begin{equation}\label{Hf}
  H_f^i(E_i) = E_i-E_{i-1}\, .
\end{equation}
In the following we will derive all general equations for an arbitrary
potential, but we will only find solutions in this special simple
case.

When we know $H_f^i(E)$ we can find the energy which on average can be
extracted with a given $n$ and $E_{\text{max}}$:
\begin{equation}\label{W1}
  W_1 = \frac{1}{Z}\sum_j\int_{E_{j-1}}^{E_j} dE\,  g(E)e^{-\beta E}\left[E-H_f^j(E)\right]
\end{equation}
where 
\[
 Z = \int_0^\infty dE\,  g(E)e^{-\beta E}
\]
is the partition function. 
If $P_i$ is the probability that the energy is
in interval $X_i$, we have
\[
 P_i = \frac{1}{Z}\int_{E_{i-1}}^{E_i}dE\, g(E)e^{-\beta E}\, .
\]
Here $i=1\cdots n+2$ where we identify $E_{n+1}=E_{\text{max}}$ and
$E_{n+2}=\infty$. That is, $P_{n+2}$ is the probability that $E>E_{\text{max}}$
and we assume that the device will not operate in this case.  The
information obtained during a measurement is on average
\[
 S = -\sum_{i=1}^{n+2} P_i\ln P_i \, .
\]
If the information is to be erased at a heat bath of temperature
$T_E$, the corresponding work of erasure is $W_{\text{er}} =
T_ES$, and Eq.~\eqref{work} is valid when $T_E=T$.

\section{How efficiently can we extract work?}

Let us assume that the maximal energy $E_{\text{max}}$ is fixed and
represents the upper limit of what our device can operate on. If the
energy is found to be above this value we can not extract it. If the
density of states does not grow too quickly, the probability of this
happening decreases quickly with increasing $E_{\text{max}}$. The free
parameters of the model are then the number $n$ of energy intervals
and the positions $E_i$ of the interval boundaries.  There are several
ways one can consider to optimize these. The simplest is to find the
maximal amount of energy $W_1$ which can be extracted in a single run
of the cycle presented in Sec.~\ref{sec:intro}, and is analyzed in
Sec.~\ref{max}. This means that we are disregarding the work of
erasure, $W_{\text{er}}$. We can also define the useful work
$W=W_1-W_{\text{er}}$, which according to Eq.~\eqref{work} is negative
when $T_E=T$. In Sec. \ref{optimal} we discuss how to maximize the
useful work or minimize the information needed at a specified
extracted work $W_1$. Finally, we can consider erasing the information
at a temperature $T_E<T$, in which case the device will operate as a
heat engine between the two thermal baths. One can then define the
efficiency $\eta=W/W_1$ in the ordinary way as the ratio of the useful
work to the energy extracted from the thermal bath. In Sec.~\ref{HE}
we will show that one can get the efficiency arbitrarily close to one
but only in the limit where $W_1 = W_{\text{er}}=0$, that is by doing
nothing, and we will find the optimal efficiency at at given $W_1$.

\subsection{Maximal $W_{1}$}\label{max}

How much energy can on average be extracted with a given $n$ and
$E_{\text{max}}$? To find this we have to maximize the work $W_1$ given by
Eq.~\eqref{W1} as a function of the energies $E_i$ marking the
boundaries of the energy intervals. In Appendix \ref{A1} it is shown
that if we consider the simplest case of $U(q)\sim q^2$ the energy
intervals $u_i=\beta(E_i-E_{i-1})$ have to satisfy the equation
\begin{equation}\label{u}
 u_{i} = 1-e^{-u_{i+1}}\, .
\end{equation}
This equation can be solved numerically, but in Appendix \ref{A1} it
is also shown how to derive an approximate solution in the limit of
large $n$. The result is that for $\beta E_{\text{max}}\gg1$ we have 
\begin{equation}\label{E}
 E_i \approx -2T\ln\left(1-\frac{i}{n}\right) \, .
\end{equation}
Using this we find 
\begin{equation}\label{WexM}
  W_1 \approx  T\left(1-\frac{2}{n}\right)
\end{equation}
\begin{equation}\label{S}
 S \approx \ln n-\ln 2 +\frac{1}{2}
\end{equation}
which shows that in the limit of large $n$ we get $W_1$ close to the
average internal energy of $T$ as given by the equipartition theorem, 
and that $W_{\text{er}}$ will grow logaritmically with $n$.

\subsection{Maximal $W$ for a given  $W_{1}$}\label{optimal}

\begin{center}
\begin{figure*}
\includegraphics[width=\linewidth]{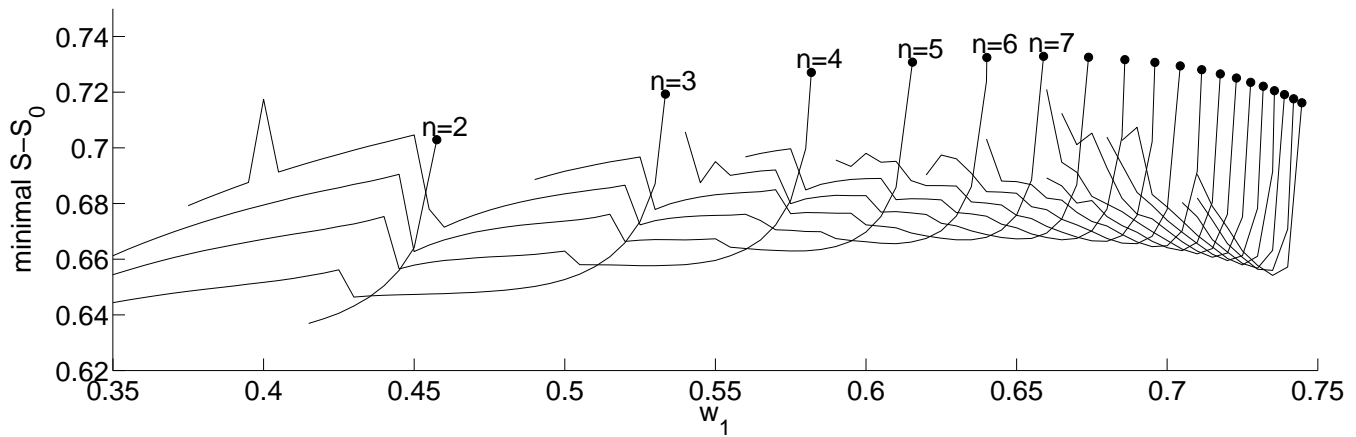}
\caption{\label{fig:S}The minimal information $S$ as a function of the extracted work $w_1$ for different $n$ and with $\beta E_{\text{max}}=3$. }
\end{figure*}
\end{center}

In the previous section we found the position of the $E_i$ such that
the extracted work, $W_1$, was maximized for a given $E_{\text{max}}$ and
$n$. However, to extract this energy we have to obtain a certain
amount of information which later has to be deleted. It is therefore
possible that by extracting an energy $W_1<W_1^{\text{max}}$ a bit less
than the maximal found above, we can reduce the information and therby
the work of erasure, $W_{\text{er}}$, so that the amount of useful work
$W=W_1-W_{\text{er}}$ could be increased. This means that we should look
for other values of $E_i$ which minimzes the information for a given
$W_1$.  Introducing the Lagrange multiplier $\lambda$ we define the function 
\begin{equation}\label{I}
 I = S+\lambda W_1 \, .
\end{equation}
We have to minimize this function subject to the constraint constraint
\eqref{W1} with a specified $W_1$. 
In Appendix~\ref{A2} it is shown
that this leads to the equations
\begin{equation}\label{eq:u2}
  \lambda(u_i + e^{-u_{i+1}}-1) 
  = \ln\frac{1-e^{-u_{i+1}}}{e^{u_{i}}-1} , \qquad i = 1\cdots n
\end{equation}
and the constraint
\begin{equation}\label{eq:c2}
 \sum_{i=1}^n v_i(e^{-v_i}-e^{v_{i+1}})=w_1 
\end{equation}
where $v_i = \sum_{j=1}^{i}u_j$ and $w_1 = \beta W_1$. We have solved
these equations numerically, using Newton's iterative
method. Numerically solving these equations is complicated by the fact
that there are in general several solutions. By choosing a large
number of initial guesses for the solution we can by reasonable
security find the solutions with the smallest necessary information,
$S$.  To make the structure clearer, it is instructive to use the not
too large value $\beta E_{\text{max}}=3$ for the maximal energy. It is also
helpful to subtract the expected  entropy, $S_0$,  according to
Eq. \eqref{S} as explained in Eq. \eqref{eq:S0}. The result is shown in
Fig.~\ref{fig:S}.
The figure shows $S-S_0$ for the minimal $S$ as a function of $W_1$
for different $n$. The dots mark the maximal $W_1$ and
corresponding $S$ for each $n$.
We can observe the following: Starting from one of the
dots of maximal $W_1$ for a given $n$, we can see that the
reduction in $S$ that can be achieved by increasing $n$ at the same
$W_1$ remains close to constant, at least for the range of $n$
studied. Also, following each curve from the dot, we see that as it
crosses the curves for larger $n$, those curves makes a jump. For
example, the $n=2$ curve crosses the $n=3$ curve around $W_1=0.43$,
and the $n=3$ curve jumps at that point. This is because as $W_1$
is reduced, the $n=3$ minimal solution has one $u_i$ which vanishes at
that point. For smaller $W_1$ this solution is not found (it is at
the boundary of the domain, and not at an interior point), while the
$n=2$ solution represents the true minimum.

\subsection{Maximal $W$ at given 
temperatures of the baths of energy and erasure}\label{HE}

\begin{center}
\begin{figure}
\includegraphics[width=\linewidth]{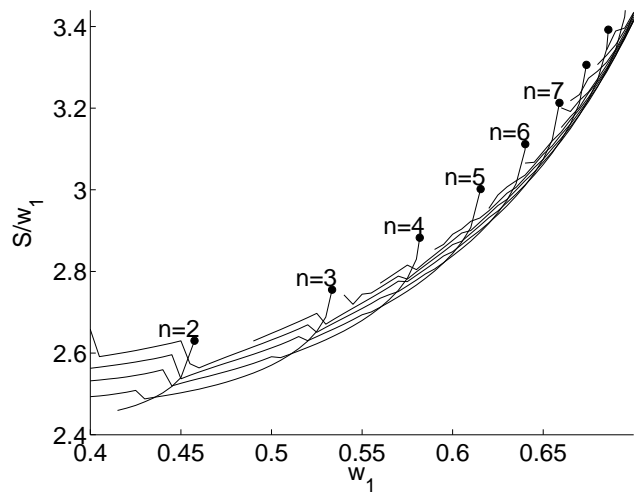}
\caption{\label{fig:sow} The ratio $S/w_1$ as a function of the extracted work $w_1$ for different $n$ and with $\beta E_{\text{max}}=3$. }
\end{figure}
\end{center}

Using the same data we can also demonstrate that there does not exist
any optimal efficiency except the trivial solution of doing
nothing, as discussed above. 
The efficiency is 
\[
 \eta = \frac{W}{W_1} = 1-\frac{T_E}{T}\frac{S}{w_1}
\]
which means that we can make the efficiency higher by making the ratio $S/w_1$
 smaller. In Fig.~\ref{fig:sow} we plot $S/w_1$
as a function of $w_1$
As we see, $S/w_1$ grows with $w_1$, which means that
we can always increase the efficiency by reducing $W_1$. For a
given $n$ this continues until one of the intervals $u_i$ (the one at
the upper limit of the energy range) collapses to zero and it becomes
favorable to decrease $n$ by one. Then the process continues with the
new $n$ until all $u_i$ are zero except $u_1$ which then cover the
whole range $[0,u_m]$.

Using the minimal $S$ as a function of $W_1$, we can also find the
maximal work $W$ which can be performed at a given temperature $T$ of
the system and when the information is erased at a lower temperature
$T_E$, and the optimal number $n_{\text{opt}}$ of energy intervals
that one needs to achieve this maximal work.
\begin{center}
\begin{figure}
\includegraphics[width=0.49\linewidth]{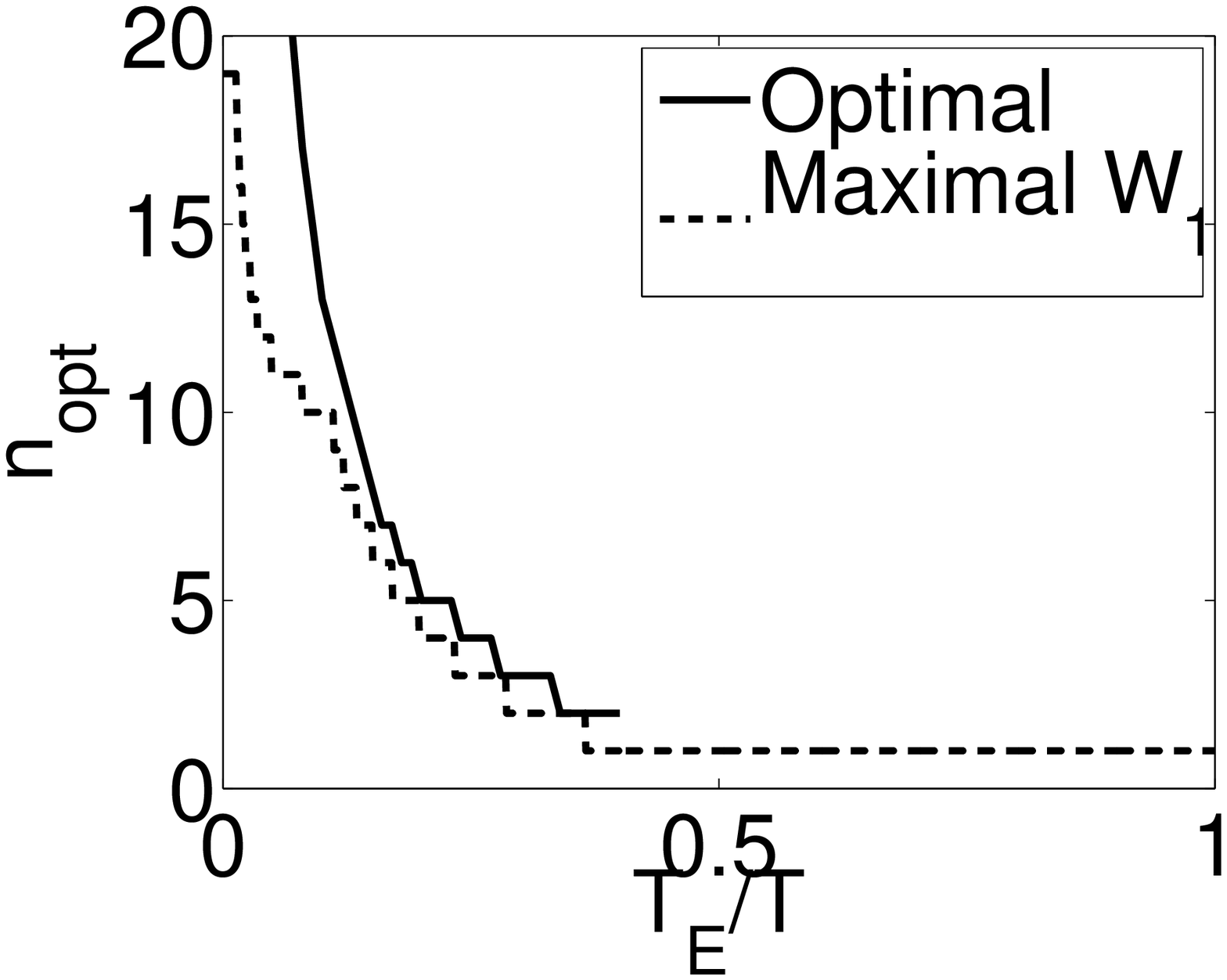}
\includegraphics[width=0.49\linewidth]{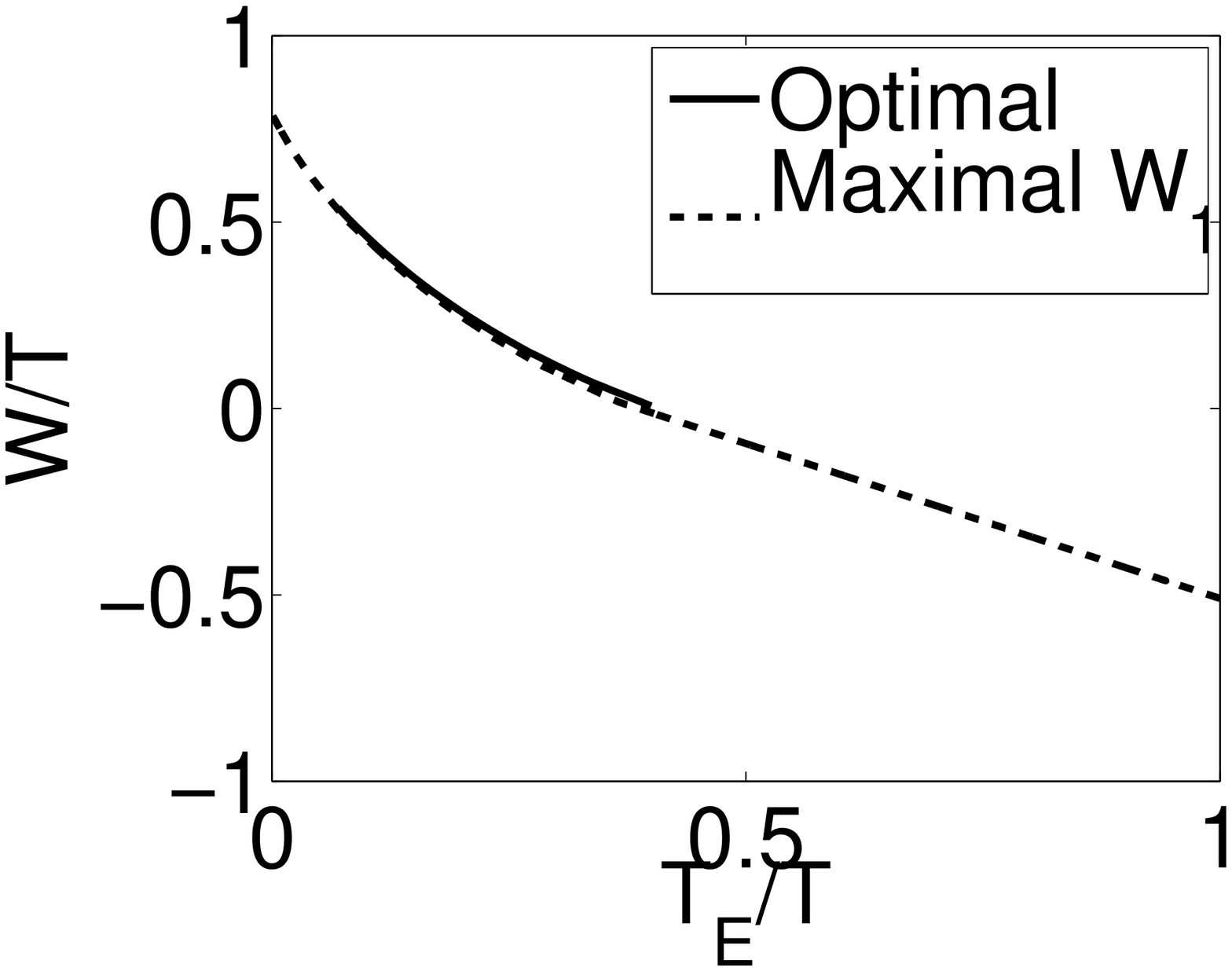}
\caption{\label{fig:maxW} $n_{\text{opt}}$ (left) and $W/T$ (right) as
  functions of $T_E/T$. In both cases two curves are shown: One based
  on the optimal solution for producing the least entropy as found in
  Sec.~\ref{optimal} (solid line) and one based on the solution giving
  the maximal heat transfer $W_1$ from the heat bath with the given $n$
  as found in Sec.~\ref{max} (dashed line). All for $\beta E_{\text{max}}=3$. }
\end{figure}
\end{center}
Figure \ref{fig:maxW} (left) shows $n_{\text{opt}}$ as a function of
$T_E/T$. Two graphs are shown, one which uses the optimal solution for
producing the least entropy as found in Sec. \ref{optimal}. The
other uses the solution giving the maximal heat transfer $W_1$ from the
heat bath with the given $n$ as found in Sec. \ref{max}. The
corresponding useful work $W/T$ is shown in Fig.~\ref{fig:maxW} (right).
As we can see, the optimal number of intervals $n_{\text{opt}}$ grows
as $T_E/T$ decreases. This is natural, since the cost of deleting
information becomes less in this case. The optimal solution has a
larger $n_{\text{opt}}$ as we expect from Fig.~\ref{fig:S} since for a
given $W_1$ we can reduce the information in measurement by increasing
the number of intervals $n$. The useful work $W$ is also larger, but
the gain in $W$ is not large and becomes smaller as $T_E/T$ decreases.

\section{Restoring reversibility: Utilizing all the information}\label{reversible}

The process discussed leads to a curious situation. We have a system
in thermal equilibrium. Then we decouple it from the environment and
make a measurement on it. That is, we gain information about the
system (the energy interval which it is in), thereby increasing our
knowledge and reducing the entropy of the system accordingly. In
principle, if the measurement process is dissipation-free, this
process is reversible, and the amount of information gained is equal
to the reduction in entropy. Then we manipulate the system in a
deterministic way, extracting energy. During this process it is
assumed that the system remains isolated from the
environment. Therefore the entropy is constant by Liouville's
theorem. The information is then erased, and this is also a reversible
process in the sense that the energy that needs to be dissipated as
heat increases the entropy of the environment by exactly the same
amount as the infomation which is deleted. The whole process is then
completely reversible. Yet, if the information is deleted at the same
temperature as the system had initially, we have seen that there will
be a net conversion of energy from mechanical energy to thermal
energy: $W_{\text{er}} > W_1$. How can this be? The answer is that the
system initially was in thermodynamic equilibrium, but after the process it is
not. This means that we have not utilized all the information that we
gained during the measurement. We have only extracted as much as
possible of the energy that was stored in the system at the moment it
was decoupled from the reservior. At the end, we are left with a
system that has lower entropy than when it started. It means that it
is a resource for extracting energy from a thermal reservoir, if it
can be reconnected to one. In order to fully exploit the information
that we gained during measurement, the system has to be returned
reversibly to its initial thermal state 
\textit{before}
 we delete the
information. If we do not do this but rather delete the information and
reconnect the system to the bath directly, this will be an
irreversible process, and this is where entropy is
generated. Returning to the steps in the process as described in
Sec.~\ref{sec:intro} we see that it is in going from step (3) and back to
step (1) that the irreversible process takes place. We now describe
how to add two further steps to the process, so that the whole cycle
becomes reversible and equality $W_{\text{er}} = W_{\text{ex}}$ of the work
of erasure and the total extracted work is restored.

\begin{enumerate}[(1)]
\item{In the initial state, the system is in thermal equilibrium with
    a bath at temperature $T$. The potential is $U(q)$ and the average
    energy is $E_1$ and the entropy $S_1$.
}
\item{We decouple the system from the bath and measure in which
    interval $X_i$ the energy lies. The average energy (which is both
    thermal average and average over measurement results) is not
    changed, $E_2 = E_1$, as it has to be since we have only measured
    and not changed the energy.  The average entropy is $S_2=S_1-S$
    where $S$ is the average information gained by the measurement.
 }
\item{We manipulate the potential in the way described in
  Sec. \ref{model}, bringing the interval $X_i$ to the bottom of the
  potential and returning the potential in the end to $U(q)$.  The
  average energy is $E_3$ and the entropy still $S_2$ since it can not
  change in an adiabatic process in an isolated system. During this
  operation the work $W_1= E_2-E_3>0$ is extracted as considered in
  previous sections. It is the maximal work which can be extracted
  keeping the system isolated.  }

\item{To reversibly return the system to the initial state we first
    modify the potential adiabatically in such a way that the
    distribution function is thermal at the right temperature
    $T$. This means that we have to find a potential $U_4(q)$ with a
    corresponding density of states $g_4(E)$ such that after the process the
    particle is left with a distribution function of the energy
    $P_4(E)$ such that
\[
 P_4(E) = \frac{1}{Z_4}e^{-\beta E} \qquad 
  Z_4   = \int_0^\infty dE \, g_4(E)e^{-\beta E} \, .
\]
Note that the potential $U_4(q)$ will in general depend on the
interval $X_i$ where the system energy was found to be. The form of
the potential can in principle be found, but we do not need it. It is
sufficient to know that it exists, which seems clear at least for
simple potentials with a single minimum.  The average energy is $E_4$
and the entropy is still $S_2$. This process requires a work $W_2 =
E_3-E_4 <0$. It is negative since the energy of the system has to
increase since we know that initially it is close to the bottom of the
potential. We have to use external work to achieve this, but it
prepares the system for the last step where a larger amount of work is
extracted from a thermal reservior.  }

\item{Finally we can now safely  reconnect the system to the bath, which is
  a reversible process and does not change anything on average, since the system
  already is prepared in a thermal state. We can then  adiabatically return the
  potential to the initial $U(q)$. This gives the same average
  energy $E_1$ and entropy $S_1$ as in the initial state. The process
  produces the work 
  \[
    W_3 = T\Delta S-\Delta U = T(S_1-S_2)-(E_1-E_4) >0\, .
  \]
}
\end{enumerate}

The total work obtained in the full cycle is 
\[
 W_{\text{ex}} = W_1+W_2+W_3  = TS
\]
which according to Landauer's principle is exactly the energy that must
be dissipated to erase the information obtained in the measurement.
This statement is true for any number of energy intervals in the
measurement scheme and any set if interval boundaries $E_i$. The only
requirement is that all processes are adiabatic, which means that they
have to be performed infinitely slowly.

\section{Summary}\label{summary}

We have discussed the model of Vaikuntanathan and Jarzynski
\cite{vaikuntanathan2011} for extracting work from a thermal bath by
measuring the energy of a particle that was thermalized with the bath
and manipulating the potential of this particle in the appropriate
way, depending on the measurement outcome. We have addressed the
question of how accurately the energy should be measured. This is
formalized in the same way as in Ref.~\cite{vaikuntanathan2011} by
dividing the energy axis in subintervals $X_i$ and assuming that the
measurement tells with perfect accuracy in which interval the energy
is. We have optimized the boundaries $E_i$ of the intervals according
to different criteria: For extracting the maximal energy, for
minimizing the entropy production and for maximizing the efficiency of
a heat engine at a given power. 

We have identified the irreversible step in the protocol of
Ref.~\cite{vaikuntanathan2011} as the one where the system is known to
be close to the lowest energy state and is reconnected with a thermal
bath. In this process the available phase space of the particle
suddenly increases, and the process is irreversible and there is a net
increase in entropy. This is in principle the same situation as in the
paradigmatic example of free expansion of an ideal gas following a
sudden increase in the accessible volume. In the context of
information driven heat engines (Maxwell's demons) similar situations
has been recently discussed. In Ref.~\cite{abreu2011} an overdamped
particle in a potential was considered and the potential was
manipulated in order to extract energy following the measurement of
position. It was found that to get the maximal work possible by the
measured information one had to strongly confine the particle
initially close to the measured position and then gradually make the
potential less steep while extracting energy. In the context of
single electron devices~\cite{horowitz2011,bergli2013} it was found
that when opening the barrier between two possible states for a
particle, this has to be done in an optimized way so that at no point
will the available phase space suddenly increase. Similarly, in this
paper we have described a protocol whereby the irreversible step in
Ref.~\cite{vaikuntanathan2011} can be reversibly performed, thereby
incresing the extracted work up to the maximal achievable by the
measured information, so that the extracted work is exactly the same
as what is needed in order to erase the information in accordance with
Landauer's principle.

\acknowledgments 

The research leading to these results has received
funding from the European Union Seventh Framework Programme
(FP7/2007-2013) under grant agreement No~308850 (INFERNOS). The author
thanks Yuri Galperin for careful reading of the manuscript.

\appendix

\section{Maximal $W_{1}$ for a given $n$}\label{A1}

We have to maximize Eq. \eqref{W1}  with respect to $E_i$:
\[
\begin{aligned}
 \pd{W_1}{E_i} = \frac{1}{Z}&\sum_jg(E_i)e^{-\beta E_i}[E_i-H_f^j(E_i)]
  (\delta_{j,i} - \delta_{j-1,i}) \\
  &- \frac{1}{Z}\sum_j\int_{E_{j-1}}^{E_j} dE\, g(E)e^{-\beta E}\pd{H_f^j(E)}{E_i}\, .
\end{aligned}
\]
Differentiating \eqref{H0} we get
\[
 \pd{H_f^j(E)}{E_i} = -\frac{g(E_j)}{g(H_f^j(E_i))}\delta_{j-1,i}\, .
\]
The equations $\frac{\partial W_1}{\partial E_i}=0$ then becomes: 
\begin{equation}\label{WexMax}
 H_f^i(E_i) 
= e^{\beta E_i}\int_{E_i}^{E_{i+1}}dEe^{-\beta E}\frac{g(E)}{g[H_f^{i+1}(E)]} \, .
\end{equation}
For $U(q) \sim q^2$  the density of states is constant, $g(E) = g_0$,
which simplifies the equation. Using Eq.~\eqref{Hf}
and 
\[
 \int_{E_i}^{E_{i+1}}dEe^{-\beta E}\frac{g(E)}{g[H_f^{i+1}(E)]}
 = -\frac{1}{\beta}\left[e^{-\beta E_{i+1}}-e^{-\beta E_{i}}\right]
\]
Eq.~\eqref{WexMax} becomes Eq.~\eqref{u}. 

We can find an approximate solution to this equation for 
 large $n$ when  all
$u_i\ll1$ and we can expand the exponential
\[
 u_{i-1} = u_i-\frac{1}{2}u_i^2+\cdots \, .
\]
Treating $i$ as a continuous variable, we get the
differential equation
\[
 \frac{du}{di} = \frac{1}{2}u^2
\]
which is integrated to give
\begin{equation}\label{uApp}
 u_i = \frac{1}{A-i/2}\, .
\end{equation}
Here $A$ is a constant of integration which has to be found from the
boundary condition $\sum_i u_i = u_m = \beta E_{\text{max}}$. We have
\[
 \int_0^n di \, u_i = -2\ln\left|\frac{n-2A}{-2A}\right| = u_m 
\]
which gives 
\begin{equation}\label{A}
 A = \frac{n/2}{1-e^{-u_m/2}}\, .
\end{equation}
We can now find 
\begin{equation}
 E_i = \frac{1}{\beta}\sum_{j<i}u_j 
  \approx \frac{1}{\beta}\int_0^i \frac{di}{A-i/2} 
= -\frac{2}{\beta}\ln\left(1-\frac{iB}{n}\right)
\end{equation}
where $B=1-e^{-u_m/2}$.

We can now calculate the extracted work and information. First we find

\[
  Z=  \int_0^\infty dE \, g_0e^{-\beta E} = \frac{g_0}{\beta}
\]
and 
\begin{equation}\label{P}
\begin{aligned}
 P_i &= \frac{1}{Z}\int_{E_{i-1}}^{E_i}dE\, g_0e^{-\beta E} 
  = e^{-\beta E_{i-1}} - e^{-\beta E_i} \\
  &= \frac{B}{n}\left[2-\frac{(2i-1)B}{n}\right] \, .
\end{aligned}
\end{equation}
The probability to find $E>E_{\text{max}}$ is $P_{n+2} = 1-e^{-\beta E_{\text{max}}}$ and the information 
\[
 S = -\sum_{i=1}^{n+2} P_i\ln P_i \, .
\]
We replace the sum by an integral:
\[
\begin{aligned}
 \int di \, & \frac{B}{n}\left[2-\frac{(2i-1)B}{n}\right] \ln \left[ \frac{B}{n}\left(2-\frac{(2i-1)B}{n}\right)\right] \\
 &= \left[B(2-B)+\frac{B^2}{n}\right]\ln\frac{B}{n} \\
&\qquad  - \frac{1}{4}\left[2-\left(\frac{2n-1}{n}\right)B\right]^2
        \left[\ln\left(\frac{2n-1}{n}\right)-\frac{1}{2}\right]\\
&\qquad   +  \frac{1}{4}\left(2+\frac{B}{n}\right)^2
        \left[\ln\left(2+\frac{B}{n}\right)-\frac{1}{2}\right]\, .
\end{aligned}
\]
When $u_m\gg1$ we have $B\rightarrow1$ and $P_{n+2}\rightarrow0$. We then
get Eq.~\eqref{S}.  Combining~\eqref{W1} and \eqref{Hf}, the
extracted work is
\begin{equation}\label{Wexg0}
  W_1 = \sum_iE_iP_{i+1} \, .
\end{equation}
Using \eqref{E} and \eqref{P} we get 
\begin{equation}\label{WexTot}
\begin{aligned}
  W_1 &= -\frac{2B}{\beta n}\int_0^ndi\, \ln\left(1-\frac{B}{n}i\right)
  \left[2-\frac{B}{n}(2i+1)\right] \\
  &= T(1-B)^2[2\ln(1-B)-1] \\ & - \frac{2TB(1-B)}{n}[\ln(1-B)-1)]
     + T\left(1-\frac{2B}{n}\right).
\end{aligned}
\end{equation}
When $u_m\gg1$ we have $B\rightarrow1$ and we find Eq. \eqref{WexM}

\begin{center}
\begin{figure}
\includegraphics[width=0.49\linewidth]{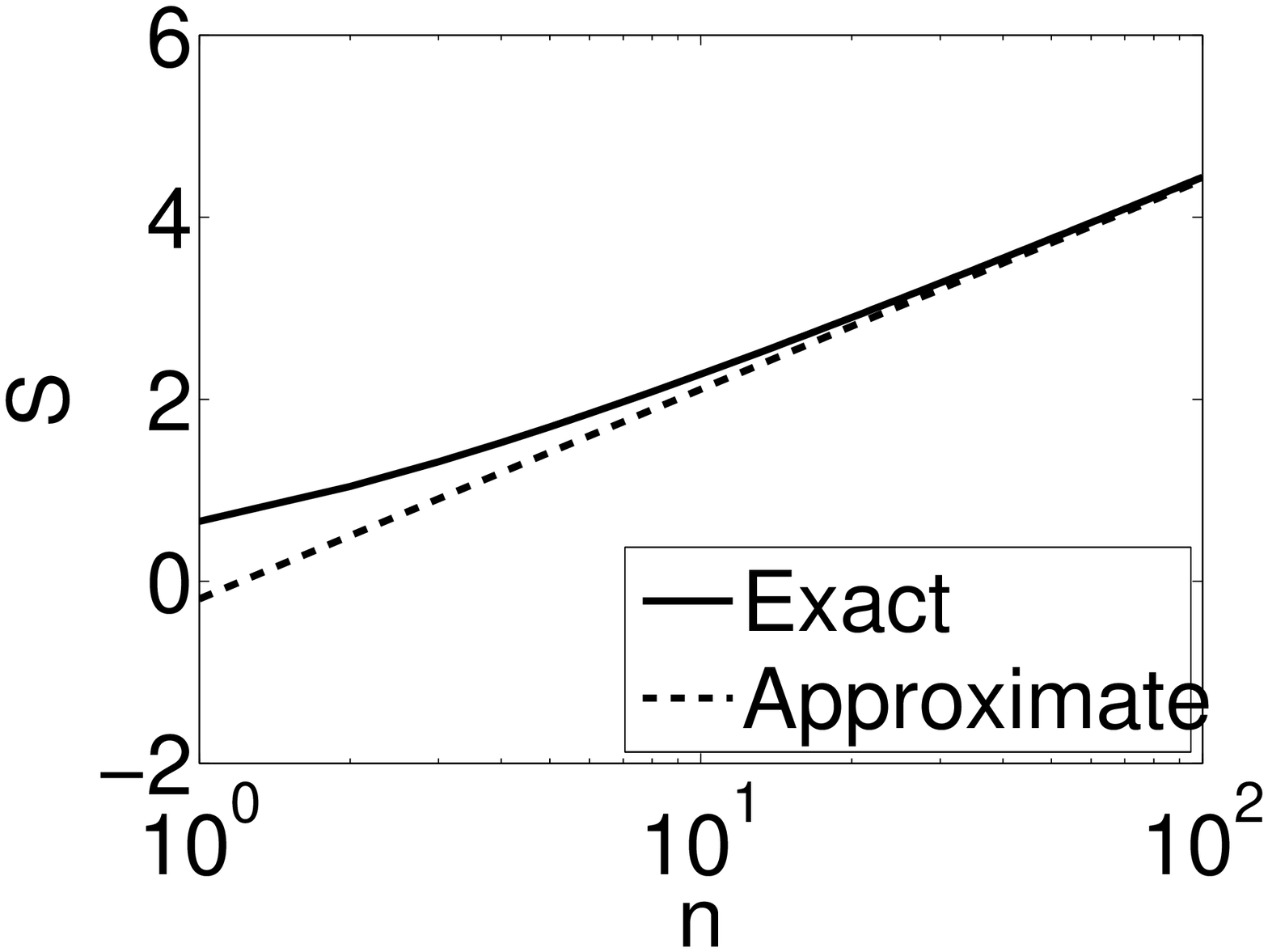}
\includegraphics[width=0.49\linewidth]{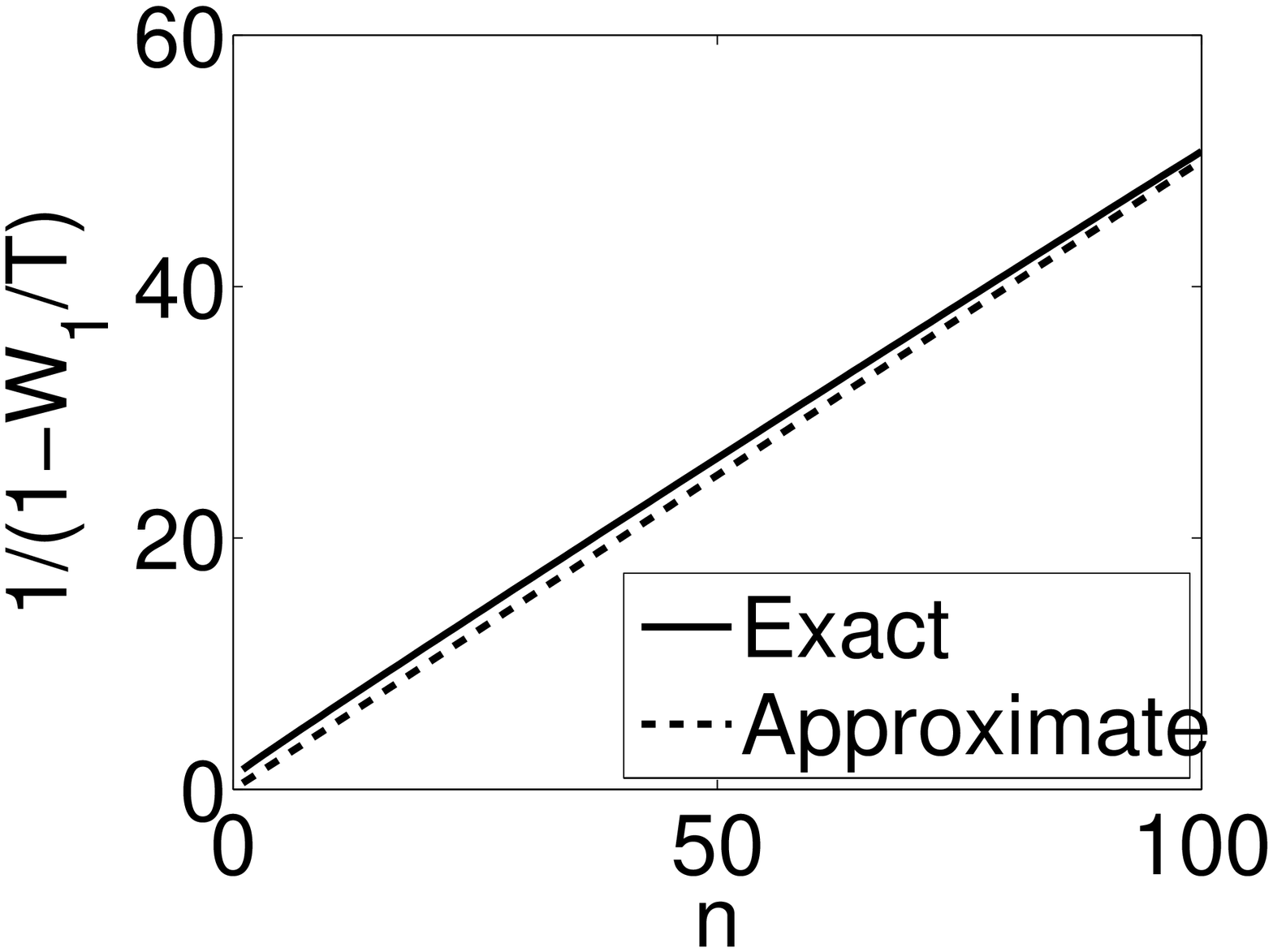}
\caption{\label{fig:n}$S$ as function of $n$ together with the
  approximate Eq. \eqref{S} (left). $1/(1-W_1/T)$ as function of $n$
  together with the approximate Eq. \eqref{WexM}. In both cases $u_m=10$. }
\end{figure}
\end{center}

To show the accuracy of the approximate solution we compare it with
the exact result found by numerical solution of Eq.~\eqref{u}.
Fig.~\ref{fig:n} (left) shows
$S$ as function of $n$ together with Eq.~\eqref{S}
while Fig.~\ref{fig:n} (right) shows $(1-W_1/T)^{-1}$ as function of $n$
together with Eq.~\eqref{WexM}, both for $u_m=10$.  We conclude that
the approximate solution works well even for $n$ not much larger than
$u_m$ which means that the $u_i$ need not be much smaller than 1.

\section{Maximal $W$ for a given  $W_{1}$}\label{A2}

To minimize $I$ in Eq.~\eqref{I} we have to solve $\partial
  I/\partial E_i=0$ together with the constraint~\eqref{W1}. We
have
\[
 \frac{\partial S}{\partial E_i} = -\sum_j (\ln P_j+1) \frac{\partial P_j}{\partial E_i} = \frac{1}{Z}g(E_i)e^{-\beta E_i}\ln\frac{P_{i+1}}{P_i}
\]
where we use 
\[
 \frac{\partial P_j}{\partial E_i} =  \frac{1}{Z}g(E_i)e^{-\beta E_i}(\delta_{j,i} - \delta_{j-1,i})
\]
and this gives
\[
 \frac{1}{\lambda}\ln\frac{P_{i+1}}{P_i} =  H_f^{i+1}(E_i) - e^{\beta E_i}\int_{E_{i}}^{E_{i+1}} dE e^{-\beta E} 
   \frac{g(E)}{g(H_f^{i+1}(E_i))}
\]
For constant $g(E)=g_0$ we get similar to \eqref{u}
\[
  \frac{1}{\lambda}\ln\frac{P_{i+1}}{P_i} =  u_{i} - 1+e^{-u_{i+1}}
\]
and from \eqref{P} we have 
\[
 P_i =  e^{-\beta E_{i-1}} - e^{-\beta E_i}
\]
which gives
\[
\frac{P_{i+1}}{P_i} = \frac{e^{-\beta E_{i}} - e^{-\beta
    E_{i+1}}}{e^{-\beta E_{i-1}} - e^{-\beta E_i}}
 = \frac{1-e^{-u_{i+1}}}{e^{u_{i}}-1}\, .
\]
The constraint is in this case given by \eqref{Wexg0} which gives
Eq. \eqref{eq:u2} and Eq. \eqref{eq:c2}.

To show the results it is instructive to subtract the expected entropy
$S_0$ according to Eq. \eqref{S} . For this, let us apply equations
\eqref{WexTot}, which we rewrite as

\[
 W_1 = C-\frac{D}{n}
\]
with
\[
\begin{aligned}
 C &= T(1-B)^2[2\ln(1-B)-1] + T\, , \\
 D &= 2BT+2BT(1-B)[\ln(1-B)-1]
\end{aligned}
\]
and \eqref{S} (it is sufficient to keep the approximate expression for
$S$, but not for $W_1$ when $E_{\text{max}}$ is not large).
Eliminating $n$ we get the relation 
\begin{equation}\label{eq:S0}
 S_0 = \ln\frac{D/2}{C-W_1}+\frac{1}{2}
\end{equation}
between the entropy $S_0$ and the extracted work. Note that this
relation is only approximate since it is based on the approximate
solution of Eq. \eqref{u}, and that Eq. \eqref{u} applies to the maximal
extracted work for a given $n$.


\begin{thebibliography}{10}
\expandafter\ifx\csname natexlab\endcsname\relax\def\natexlab#1{#1}\fi
\expandafter\ifx\csname bibnamefont\endcsname\relax
  \def\bibnamefont#1{#1}\fi
\expandafter\ifx\csname bibfnamefont\endcsname\relax
  \def\bibfnamefont#1{#1}\fi
\expandafter\ifx\csname citenamefont\endcsname\relax
  \def\citenamefont#1{#1}\fi
\expandafter\ifx\csname url\endcsname\relax
  \def\url#1{\texttt{#1}}\fi
\expandafter\ifx\csname urlprefix\endcsname\relax\def\urlprefix{URL }\fi
\providecommand{\bibinfo}[2]{#2}
\providecommand{\eprint}[2][]{\url{#2}}

\bibitem[{\citenamefont{Allahverdyan and
  Nieuwenhuizen}(2002)}]{allahverdyan2002}
\bibinfo{author}{\bibfnamefont{A.}~\bibnamefont{Allahverdyan}}
  \bibnamefont{and}
  \bibinfo{author}{\bibfnamefont{T.}~\bibnamefont{Nieuwenhuizen}},
  \bibinfo{journal}{Physica A: Statistical Mechanics and its Applications}
  \textbf{\bibinfo{volume}{305}}, \bibinfo{pages}{542 } (\bibinfo{year}{2002}),
  ISSN \bibinfo{issn}{0378-4371},
  \urlprefix\url{http://www.sciencedirect.com/science/article/pii/S03784371010%
06057}.

\bibitem[{\citenamefont{Campisi}(2008)}]{campisi2008}
\bibinfo{author}{\bibfnamefont{M.}~\bibnamefont{Campisi}},
  \bibinfo{journal}{Studies in History and Philosophy of Science Part B:
  Studies in History and Philosophy of Modern Physics}
  \textbf{\bibinfo{volume}{39}}, \bibinfo{pages}{181 } (\bibinfo{year}{2008}),
  ISSN \bibinfo{issn}{1355-2198},
  \urlprefix\url{http://www.sciencedirect.com/science/article/pii/S13552198070%
00974}.

\bibitem[{\citenamefont{Jarzynski}(1997)}]{jarzynski1997}
\bibinfo{author}{\bibfnamefont{C.}~\bibnamefont{Jarzynski}},
  \bibinfo{journal}{Phys. Rev. Lett.} \textbf{\bibinfo{volume}{78}},
  \bibinfo{pages}{2690} (\bibinfo{year}{1997}),
  \urlprefix\url{http://link.aps.org/doi/10.1103/PhysRevLett.78.2690}.

\bibitem[{\citenamefont{Vaikuntanathan and
  Jarzynski}(2011)}]{vaikuntanathan2011}
\bibinfo{author}{\bibfnamefont{S.}~\bibnamefont{Vaikuntanathan}}
  \bibnamefont{and}
  \bibinfo{author}{\bibfnamefont{C.}~\bibnamefont{Jarzynski}},
  \bibinfo{journal}{Phys. Rev. E} \textbf{\bibinfo{volume}{83}},
  \bibinfo{pages}{061120} (\bibinfo{year}{2011}),
  \urlprefix\url{http://link.aps.org/doi/10.1103/PhysRevE.83.061120}.

\bibitem[{\citenamefont{Sato}(2002)}]{sato2002}
\bibinfo{author}{\bibfnamefont{K.}~\bibnamefont{Sato}}, \bibinfo{journal}{J.
  Phys. Soc. Jpn.} \textbf{\bibinfo{volume}{71}}, \bibinfo{pages}{1065}
  (\bibinfo{year}{2002}).

\bibitem[{\citenamefont{Marathe and Parrondo}(2010)}]{marathe2010}
\bibinfo{author}{\bibfnamefont{R.}~\bibnamefont{Marathe}} \bibnamefont{and}
  \bibinfo{author}{\bibfnamefont{J.~M.~R.} \bibnamefont{Parrondo}},
  \bibinfo{journal}{Phys. Rev. Lett.} \textbf{\bibinfo{volume}{104}},
  \bibinfo{pages}{245704} (\bibinfo{year}{2010}),
  \urlprefix\url{http://link.aps.org/doi/10.1103/PhysRevLett.104.245704}.

\bibitem[{\citenamefont{Landauer}(1961)}]{landauer1961}
\bibinfo{author}{\bibfnamefont{R.}~\bibnamefont{Landauer}},
  \bibinfo{journal}{IBM J. Res. Dev.} \textbf{\bibinfo{volume}{5}},
  \bibinfo{pages}{183} (\bibinfo{year}{1961}).

\bibitem[{\citenamefont{Abreu and Seifert}(2011)}]{abreu2011}
\bibinfo{author}{\bibfnamefont{D.}~\bibnamefont{Abreu}} \bibnamefont{and}
  \bibinfo{author}{\bibfnamefont{U.}~\bibnamefont{Seifert}},
  \bibinfo{journal}{EPL (Europhysics Letters)} \textbf{\bibinfo{volume}{94}},
  \bibinfo{pages}{10001} (\bibinfo{year}{2011}),
  \urlprefix\url{http://stacks.iop.org/0295-5075/94/i=1/a=10001}.

\bibitem[{\citenamefont{Horowitz and Parrondo}(2011)}]{horowitz2011}
\bibinfo{author}{\bibfnamefont{J.~M.} \bibnamefont{Horowitz}} \bibnamefont{and}
  \bibinfo{author}{\bibfnamefont{J.~M.~R.} \bibnamefont{Parrondo}},
  \bibinfo{journal}{EPL (Europhysics Letters)} \textbf{\bibinfo{volume}{95}},
  \bibinfo{pages}{10005} (\bibinfo{year}{2011}),
  \urlprefix\url{http://stacks.iop.org/0295-5075/95/i=1/a=10005}.

\bibitem[{\citenamefont{Bergli et~al.}(2013)\citenamefont{Bergli, Galperin, and
  Kopnin}}]{bergli2013}
\bibinfo{author}{\bibfnamefont{J.}~\bibnamefont{Bergli}},
  \bibinfo{author}{\bibfnamefont{Y.~M.} \bibnamefont{Galperin}},
  \bibnamefont{and} \bibinfo{author}{\bibfnamefont{N.~B.}
  \bibnamefont{Kopnin}}, \bibinfo{howpublished}{arXiv:1306.2742}
  (\bibinfo{year}{2013}).

\end{thebibliography}
\end{document}